\documentclass[manuscript]{aastex}

\slugcomment{}

\shorttitle{The Light Curve Solutions of an Eclipsing Binary OGLE-GD-ECL-04451 with Dramatic Changes in Amplitude}
\shortauthors{Li Gang et al.}

\begin{document}


\title{Light Curve Solutions of an Eclipsing Binary OGLE-GD-ECL-04451 with a Dramatic Change in Amplitude}


\author{Li Gang, Fu Jianning\altaffilmark{1} and Zhang Yanping}
\affil{Department of Astronomy, Beijing Normal University}


\altaffiltext{1}{Send offprint request to: jnfu@bnu.edu.cn}

\label{abstract}
\begin{abstract}
We present light curve solutions of the W UMa-type eclipsing binary OGLE-GD-ECL-04451, observed by both the \emph{Optical Gravitational Lensing Ex-periment} (\emph{OGLE}) program in 2006 and the \emph{Antarctica Survey Telescope} (\emph{AST3-1}) in 2012 at Dome A. We analyzed this binary system with the Wilson-Devinney(W-D) method 2013 version and derived the mass ratio $q=2.91 \pm 0.07$, the inclination $i=76.86^\circ \pm 0.23^\circ$, and the light variattion amplitud was $0^m.51$ based on the \emph{OGLE} data. From the \emph{AST3-1}'s data, we find that the amplitude dropped to $0^m.44$(2012) and the difference of magnitudes of the two light maxima is $0^m.03$. A hot spot was then added on the surface of the secondary to demonstrate the amplitude change and O'Conell effect of the binary system. 
\end{abstract}





\section{Introduction}

In eclipsing binary system the stellar spot is an vital parameters which can cause some shape change of light curves\citep{2014AJ....147...98L,2012PASJ...64...48L,2012MNRAS.423.3646Q}. And time-evolution of spots have been also detected by Chinese Small Telescope ARray (CSTAR) during a half-year observation in Dome A, Antarctica\citep{2014ApJS..212....4Q}. The most apparent effect caused by spots is O'Connell effect, which means that two maximum values show inconsistency although one can see both of the binary in the same time\citep{1951PRCO....2...85O}. Some phase displacements in extreme values of light curve as well as O-C variation are ascribed to the exist of spots, even no O'conell effect was found\citep{2015AJ....150...70L,2013ApJ...774...81T}. What's more, \cite{2013ApJ...774...81T} gave an brief mathematical explanation about phase displacement and O-C anti-correlation.

W UMa-type binaries are attractive contact binaries because both components overfill the critical Poche lobe and share a common convective envelope. They are divided into two subtypes: A-subtype and W-subtype. A-subtype represents that the higher the temperature of a component is, the bigger the mass of this component is, while W-subtype is just the opposite. The origin and evolution of such type binaries was disucussed in several studies\citep{1981Ap&SS..78..401V}. They thought that detached binaries should evolve to this type of binary because of angular momentum loss(AML) via stellar winds.

Wilson-Devinney(W-D) code is a powerful program to deduce the parameter of an binary system\citep{1971ApJ...166..605W,1979ApJ...234.1054W,1990ApJ...356..613W,1994PASP..106..921W}. Several important parameters can be determined by it, like the inclination, temperature and mass ratio. If one combine light curve and both radial velocities of two components, the orbital semi-major axis and star radii as well as star masses can be calculated. For most cases one only have light curve data, the inclination and eccentricity can be deduced, but one can only get a mass ratio\citep{2006ASPC..349...71W}. It will be still helpful if combining other observations.

\section{Observation Data}


The star, OGLE-GD-ECL-04451, was firstly observed by the \emph{Optical Gravitational Lensing Ex-periment} (\emph{OGLE}) program in 2006\citep{2010AcA....60..295S} and then re-observed by the \emph{Antarctica Survey Telescope} (\emph{AST3-1}) at Dome A, Antarctica in 2012. Its coordinates are $\alpha = 10:53:45.48, \delta = -61:15:35.0$ in International Celestial Reference System (ICRS). The basic information is listed in Table~\ref{tab1} from UCAC4 catalog \citep{2013AJ....145...44Z}.

The OGLE program aimed to detect the optical gravitational lensing phenomenon. Tts data have good accuracy but low time resolution in general. The uncertainty reached typically $\sim 0^m.03$ for this binary star with the exposure time of $\sim1000s$. OGLE kept observing OGLE-GD-ECL-04451 in almost one year and collected 1688 frames in I band and 8 frames in V band about this target. Theephemeris was given as follows,

\begin{equation}
Min.I=2453965.13760+0^d.41731592 \times E
\end{equation}

\begin{figure}
\epsscale{.80}
\plotone{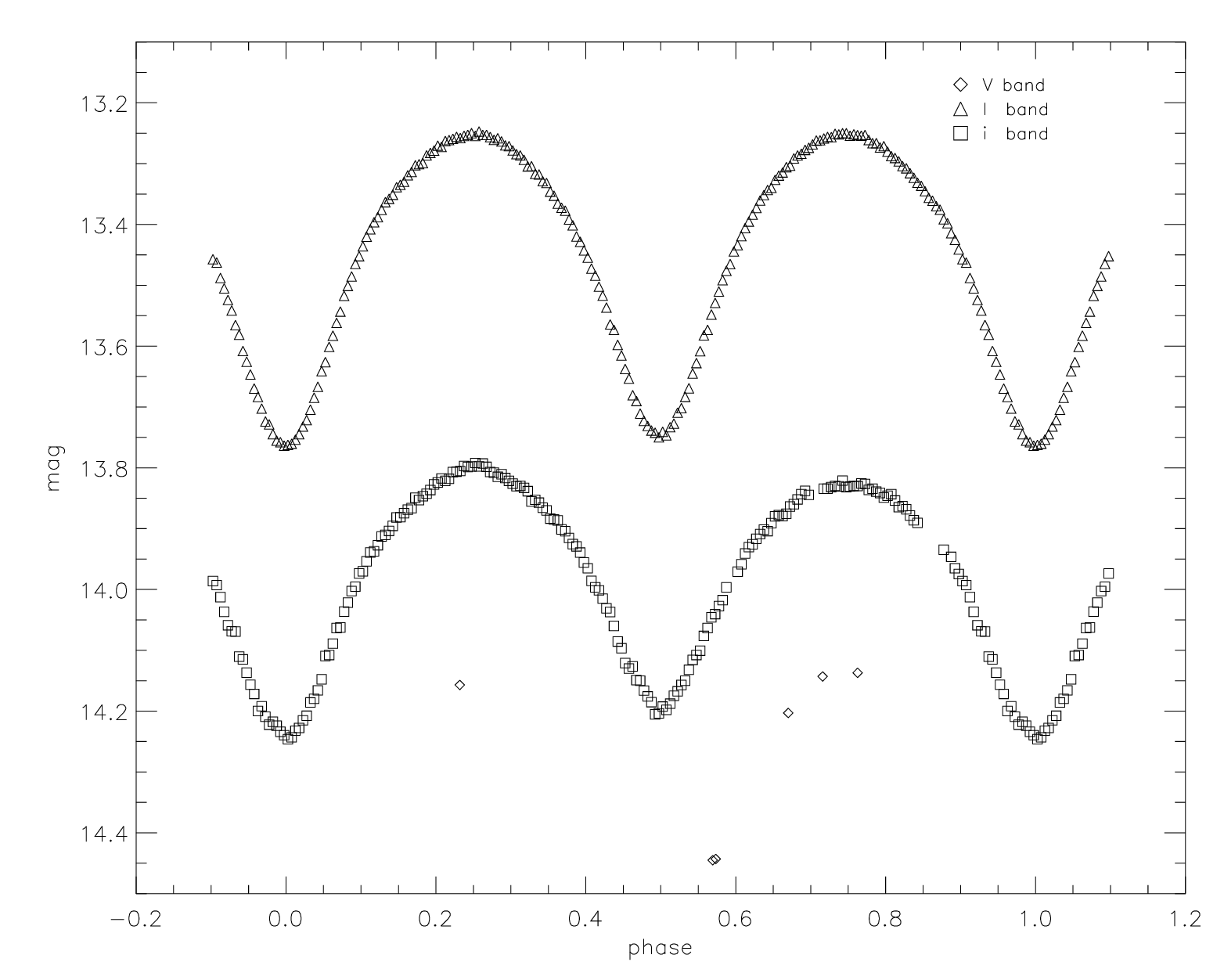}
\caption{All the data of OGLE-GD-ECL-04451. Diamond stands for the data in V from OGLE and square stands for the data in I from OGLE, while triangle stands for the data in \emph{i} from AST3-1.\label{fig1}}
\end{figure}

On the other hand, AST3-1 is a 50/68 cm Schmidt-like equatorial-mount telescope and the first trackable telescope of China in Antarctica inland. AST3-1 observed the target OGLE-GD-ECL-04451 from the $25^{th}$ to the $30^{th}$ of April in 2012 but the data collected in $27^{th}$ were discarded due to low quality. This target is one of the results in variable-star survey program made by us. There are in total 1648 frames in \emph{SDSS i} band. The data information are listed in Table~\ref{tab2}. Phased light curves are shown in Figure~\ref{fig1}.

\section{O-C analysis}
\begin{figure}
\epsscale{1}
\plotone{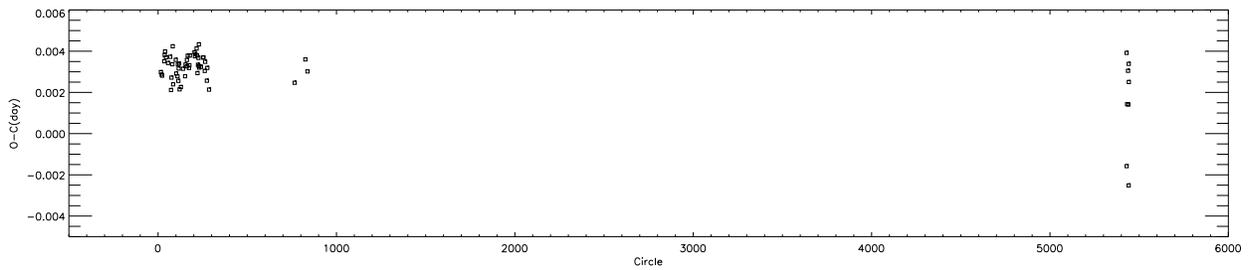}
\caption{The O-C diagram. \label{fig2}}
\end{figure}

O-C stands for O[bserved] \emph{minus} C[alculated] and is a widely used method when discussing cyclic phenomena\citep{2005ASPC..335....3S}. Figure~\ref{fig2} shows the O-C diagram of OGLE-GD-ECL-04451 and Table~\ref{tab3} lists the O-C values. We selected the data close to the calculated minimum time and used the two-order fitting to derive the Observed time, whose uncertainty were estimated by the monte-carlo method. Since the sampling interval of the \emph{OGLE} data is large, there are in total 62 minimum epoches calculated for the one-year observation data. In the meanwhile, 8 minimum times are determined from the \emph{AST3-1} data. One can see that there is not apparent oscillation or variation trend in the O-C diagram, showing the stability of the binary's system. However, due to the exist of a big gap in the O-C diagram, more data are needed for a more convincing conclusion.

A parabolic fitting is used to estimate the upper limit of period variation rate ($\mathrm{d} P / \mathrm{d} t$). Since the variation of period $P(t)$ is extremely small, only the first order of period variation is assumed as follows:

\begin{equation}
P(t)=P_0+\frac{\mathrm{d} P}{\mathrm{d} t} t
\end{equation}

considering that the number of circles $E=t / \overline{P}$  and integrating with respect of E, one gets:

\begin{equation}
T_m = T_0 + P_0 E+\frac{1}{2}\frac{\mathrm{d} P}{\mathrm{d} t} \overline{P} E^2
\end{equation}

where $T_m$ means a minimum time and $T_0$ means the first minimum time. A new ephemeris is,

\begin{equation}
T_m = 2453777.3450(6) + 0.417315(3) E - 3(3)\times 10^{-11} E^2
\end{equation}

From equation (4) one can see that $P_0=0.417315(3)$ days and $\mathrm{d} P / \mathrm{d} t=1.4(1.4) \times 10^{-10}$. So It is hard to make sure that the $\mathrm{d}P/ \mathrm{d} t$ is detected.

\section{Light Curve Solutions by the W-D Method}
The light curves were converted to phased light curves with the period value given in equation (4). All of the photometric observations in band I and \emph{i} were averaged to 200 normal data points in order to improve the speed of program. The W-D program (2013 version) was used to analyze the light curves. We noted that the shape in the different two years changed a lot: (1) data in I band collected in 2006 showed no O'Connell effect with amplitude of $0^m.51$; (2) data in \emph{i} band collected in 2012 showed an apparent O'Connel effect with the maxima difference of $0^m.03$ and amplitude dropped to $0^m.44$; (3) the number of points in V band collected in 2006 is only 8, which is not suitable for analysis. In short, we selected the data in I in 2006 and in \emph{i} in 2012 to look for solutions, respectively.

The choice of initial values of parameters is vital in W-D program iteration, especially the effective temperature of the primary component($T_1$). Luckily, the photometric data in I band and V band gave the colour index to deduce $T_1$, which at phase=0.75 is $V-I=0^m.90$, corresponding to $5628K$ hence a G5 type star \citep{1966ARA&A...4..193J}. As the colour index is only a rough estimate of the temperature, we selected $T_1$=5550K, 5600K, 5660K and 5710K as initial values and derived 4 groups of solution in each bands. The gravity-darkening coefficients g1=g2=0.32 \citep{1976ApJ...205..208L} and bolometric albedos A1=A2=0.5 \citep{1969AcA....19..245R}. In the 2013 version of W-D program, the limb-darkening coefficients can be calculated automatically based on the method from \cite{1993AJ....106.2096V}. We selected mode 3 for overcontact binaries, in which the inclination \emph{i}, the effective temperature of the secondary component $T_2$, the modified surface potential $\Omega_1=\Omega_2$, the mass ratio $q=M_2/M_1$ and the luminosity of the primary component $L_1$ can be adjusted.

\subsection{Solutions with Light Curve in 2006}

\begin{figure}
\epsscale{1}
\plotone{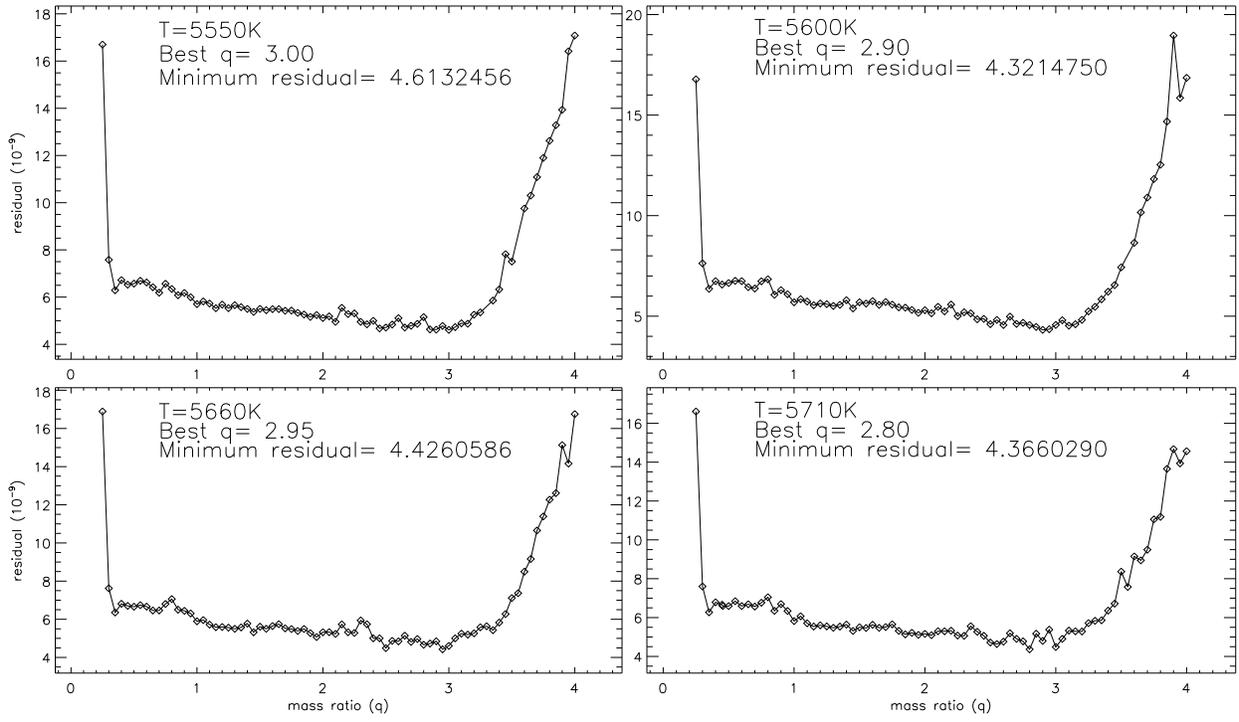}
\caption{The residual-q diagram. \label{fig3}}
\end{figure}

Due to the lack of previous research of OGLE-GD-ECL-04451, the q-search method was firstly used to look for the best mass ratio \citep{1996A&AS..117..105N}. We fixed the q value and adjusted other 4 parameters to a converged result. Then, $T_2$ was set as the same as $T_1$ and $\Omega_1=\Omega_2$ can be derived from table \emph{critout.tab} in W-D program's help file. $L_1$ is set to $0.3\times 10^{-5}$ because the magnitude zero corresponds to the observable light of 1.0000\footnote{Thanks for professor R.E.Wilson's nice help}. The q range is from 0.25 to 4 and the step is set as 0.05. Finally 302 groups of results in 4 $T_1$ were obtained and one can see in Figure~\ref{fig3} that the best q value is $\sim2.9$. However, slight discrepancy can be seen with different $T_1$. These q values are set as the initial values in 4 cases.

\begin{figure}
\epsscale{1}
\plotone{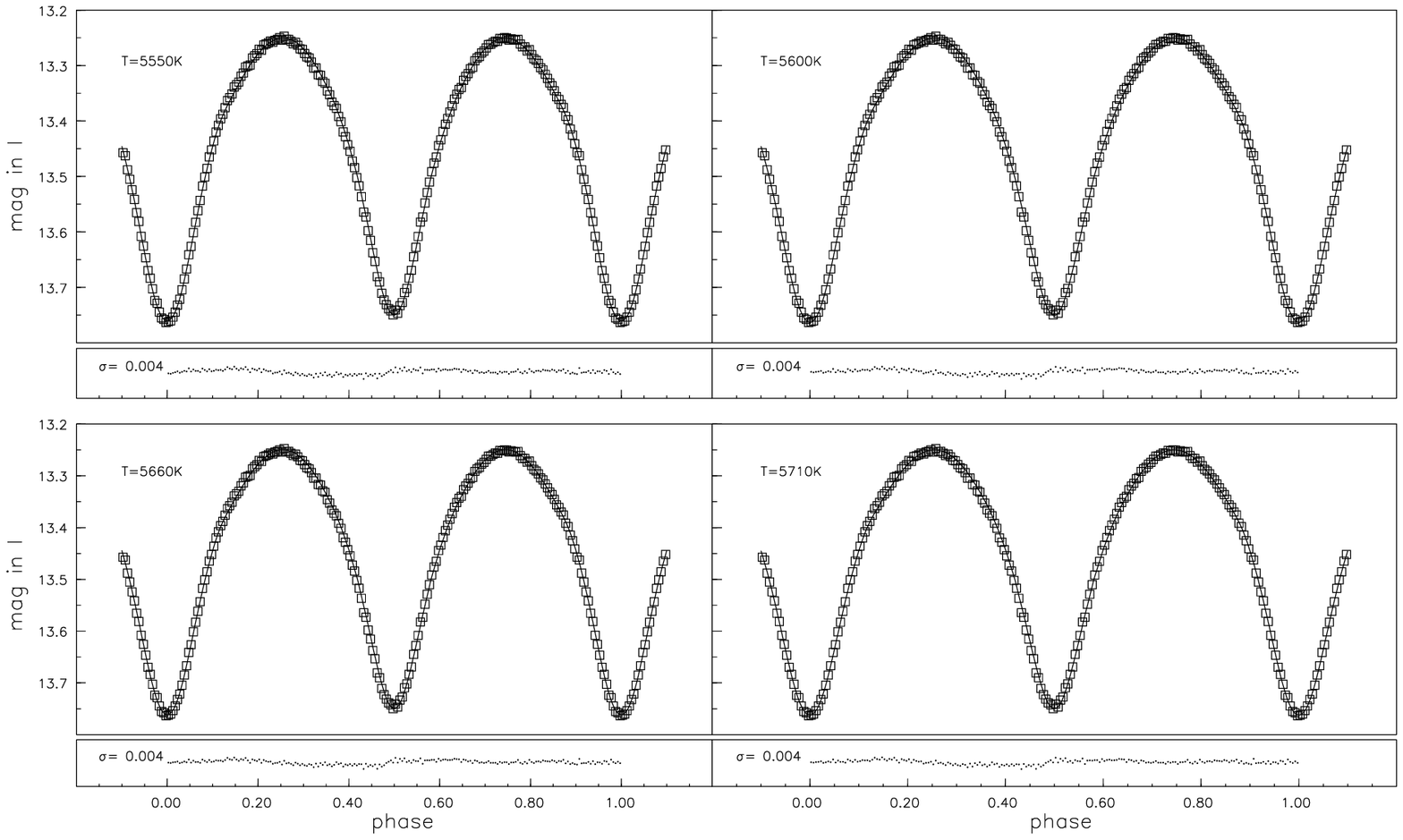}
\caption{The results in I in 2006. The square stands for the observed data and solid line stands for theoretical light curves, while the dots represent to the residual. \label{fig4}}
\end{figure}

As the second step, we started iteration from the results whose q is the best. We list the results in 4 $T_1$ based on the data in I band in 2006 at $2^{rd}$, $4^{th}$, $6^{th}$ and $8^{th}$ columns in Table~\ref{tab4} and one can see the best fitting results in Figure~\ref{fig4}.

\subsection{The Spot in 2012}

As indicated in Section 4, the shape change and the O'Connell effect appeared in the light curves from AST3-1 in 2012. So, spot models were adopted to fit the observed light curves. It is common because both components are fast-rotating solar-type stars so they would show solar-like activity, including photospheric spots\citep{2015AJ....149...38Q}. The well accepted explanation of the systems with spots was proposed by \cite{1975ApJ...198..563M}, who indicated that it is more possible to locate a spot on the more massive component. In order to decrease both of the maximum magnitudes simultaneously, we tried a number of solutions and finally found that there should be a big spot in the polar regions. 

\begin{figure}
\epsscale{1}
\plotone{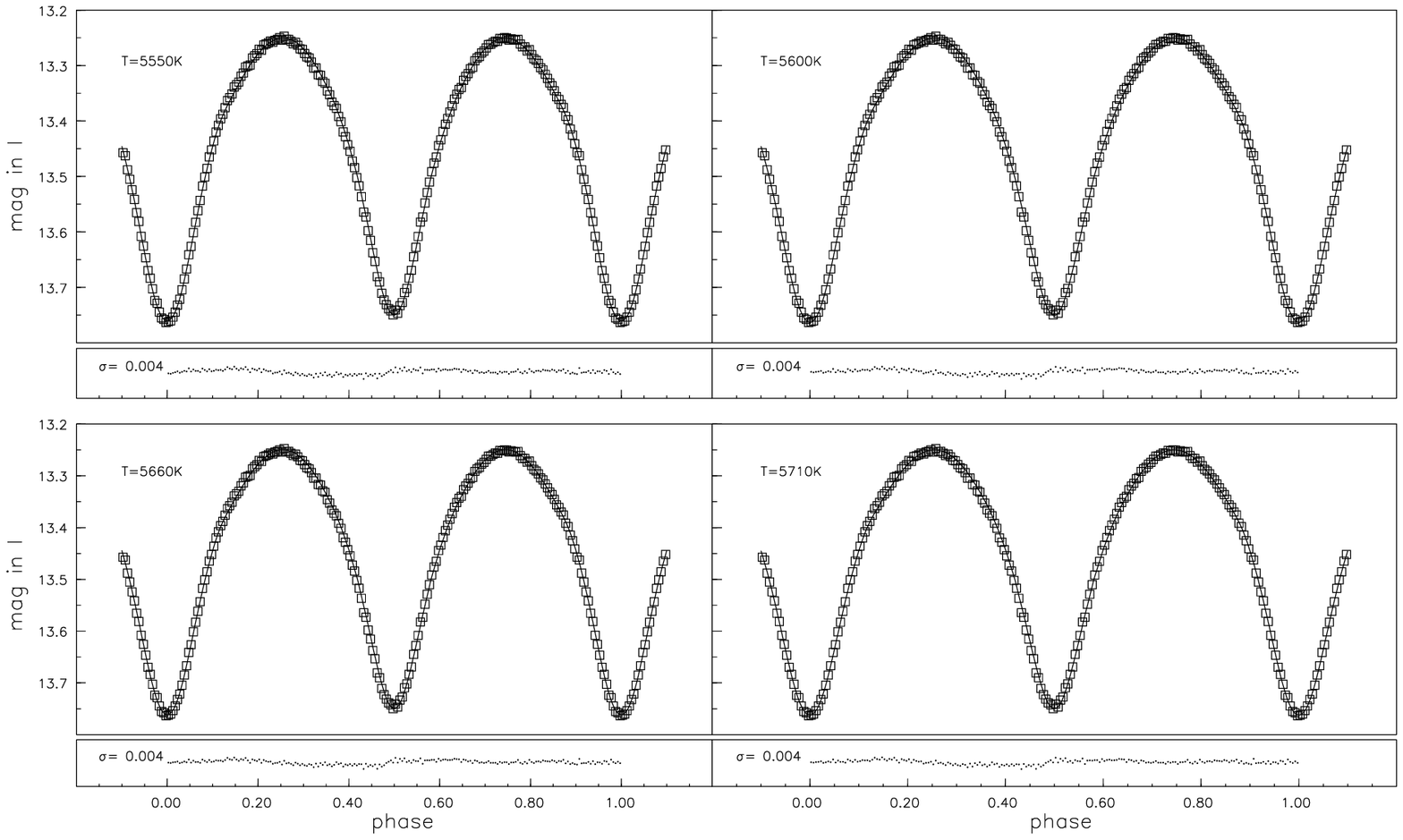}
\caption{The results in \emph{i} in 2012. The square stands for the observed data and solid line stands for theoretical light curves, while the dots represent to the residual. \label{fig5}}
\end{figure}

\begin{figure}
\epsscale{1}
\plotone{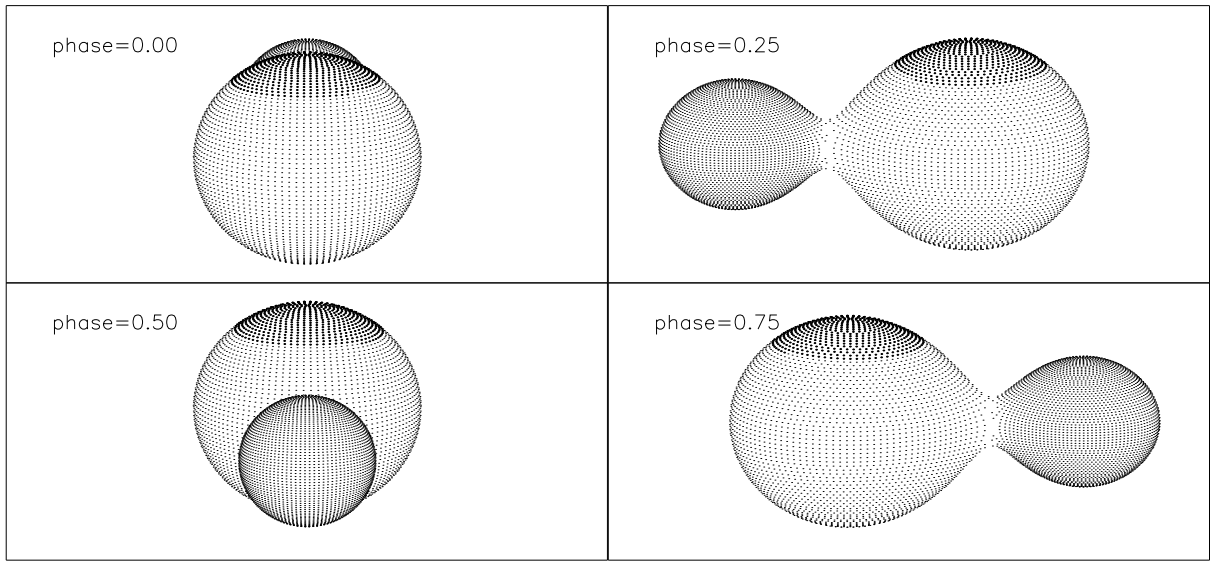}
\caption{The geometric structure of this binary system. Note that a big spot is placed at north pole of secondary component. \label{fig6}}
\end{figure}

Due to the conservation of angular momentum and low mass transfer rate from the stable O-C result, we assumed that the inclination \emph{i} and the mass ratio q would not change dramatically during the six-year evolution of the binary system from 2006 to 2012. We hence adjusted the effective temperature of the secondary component $T_2$, the modified surface potential $\Omega_1=\Omega_2$ and the luminosity of the primary component $L_1$ in \emph{i} band. As for the spot, the latitude $\theta$, the longtitude $\phi$, the angular radius $r_{spot}$ as well as the temperature factor $T_s/T_*$ (which means the ratio between the temperature of the spot and the temperature of the star) are adjustable. One can see the results listed at $3^{rd}$, $5^{th}$, $7^{th}$ and $9^{th}$ columns in Table~\ref{tab4} with all the parameters converged. Figure~\ref{fig5} shows the theoretical light curves in \emph{i} band while Figure~\ref{fig6} shows the geometric structure of this binary system, where one can see the effect of one spot in both of the two maximum times.

\section{Conclusions and Discussions}

In this paper we researched an eclipsing binary OGLE-GD-ECL-04451. O-C diagram shows no variation of this system and light curve solutions were obtained by W-D method. Firstly, we estimated $T_1 \approx 5628K$ by colour index $V-I=0^m.90$ from OGLE data in 2006 and decided four initial values to estimate the uncertainty caused by $T_1$. In next step, light curve solutions based on the data in I observed in 2006 were got. The mass ratio $q=2.91 \pm 0.07$ and inclination $i=76.86^\circ \pm 0.23^\circ$, which show that it is a W-subtype W UMa-type binary. Then, amplitude decrease and O'Connell effect were found in the data observed in band \emph{i} in 2012. So a big and hot spot was adopted to fit the light curve better. 

\subsection{Reliability of the Results}

Since we can not determine $T_1$ directly and just make an estimate by the colour index of whole system, maybe there will be some inconsistancy between different groups of solution based on different $T_1$ we assumed. However, the solusions in four $T_1$ values reveal prefect astringency, as all the parameters ($T_2$ is replaced by $\Delta T=T_1-T_2$) have similar values after iteration. At $10^{th}$ and $11^{th}$ columns in table~\ref{tab4}, we give an overall parameters, which is averaged from previous solutions. One can see that $i=76.86^{\circ} \pm 0.23 ^{\circ}$ and $q=2.91\pm 0.07$. q greater than 1 stands for a W-subtype binary.

Another item is about spot. Was the light curve in 2006 contaminated by spot? Assume that if we hadn't get the data in 2006, a smaller amplitude in 2012 would have been adopted when getting solutions and the results might have changed a lot. So it should be cautious that the results may be influenced by a spot in 2006 which changed the amplitude but didn't cause O'Connell effect. 

\subsection{Spot Activity}
Spot activity has been detected for a long time. \cite{1947PASP...59..261K} found unaccountable fluctuation in eclipsing binary AR Lac, which may be caused by spot activity. After that, stellar spot activity has been proved for numerous observation. Similar to solar spot, stellar spot is believed to be the result of stellar magnetic field. But comparing to our sun, different stars have various properties about spot distribution and activity. There is evidence that late-type stars with fast rotation can have huge spots' distribution in polar region, rather than small spots in equator like sun. For example, \cite{2009A&ARv..17..251S} displayed the spot distribution of a young solar-like star \emph{EK Draconis} by doppler image technology, whose huge spots are located in polar region. Also \cite{1995ApJ...452..879N} found that more than 50\% area of an overcontact binary II Peg was covered by cool spots. Some empirical relationships between spots' properties and stellar parameters were derived by \cite{2005LRSP....2....8B,1996ApJ...463..766O,2008A&A...479..557F,2005MNRAS.356.1501B}.

In terms of our binary system OGLE-GD-ECL-04451, It converged to a same result from different $T_1$ that there is a big (occupied $13.4\% \pm 1.7\%$ area of stellar surface) but hot (the temperature factor is $1.63\pm 0.21$) spot postulated to be placed at north pole in secondary component. Although such a big spot distribution has been proved as we mentioned in last paragraph, the higher temperature is hard to be explained by magnetic activity. Generally, mass transfer is adopted to explain hot spot, which can cause both hot spot near the neck region of the common envelope and period change in O-C diagram. However, these two consequense didn't be confirmed from our results.




\acknowledgments
The authors thank the support from the Scientific Research and Entrepreneurship Plan of Beijing College Students. JNF acknowledges the support from the Joint Fund of Astronomy of National Natural Science Foundation of China (NSFC) and Chinese Academy of Sciences through the Grant U1231202, and the support from the National Basic Research Program of China (973 Program 2014CB845700 and 2013CB834900). Also, We are grateful to R. E. Wilson for guiding our W-D program.

\clearpage



\clearpage

\begin{table}
\begin{center}
\caption{Information of OGLE-GD-ECL-04451 from UCAC4 Catalog.\label{tab1}}
\begin{tabular}{rrrrrrrrrr}
\hline\hline
$\alpha$ (ICRS) & $\delta$ (ICRS) & pmRA & pmDE & Jmag & Kmag & Bmag & Vmag & rmag & imag\\
\hline
163.4394856 & -61.2597137 & -7.7 & 1.2 & 12.428 & 11.998 & 15.112 & 14.148 & 13.899 & 13.893\\
\hline
\end{tabular}
\end{center}
\end{table}

\begin{table}
\begin{center}
\caption{Data Information\label{tab2}}
\begin{tabular}{rrrrrr}
\hline\hline
telescope & band & frame & expo time & duration & accuracy\\
\hline

OGLE in 2006 & I & 1688 & $\sim$1000s & 364 days & $0^m.003$\\
OGLE in 2006 & V & 8 & $\sim$1400s & 77 days & $0^m.003$\\
AST3-1 in 2012& i & 1675 & 60s & 5 days & $0^m.016$\\
\hline
\end{tabular}
\end{center}
\end{table}

\begin{table}
\begin{center}
\tiny
\caption{O-C Results. The ephemeris is $Min.I=2453965.13760+0^d.41731592 \times E$ from OGLE program. \label{tab3}}
\begin{tabular}{rrrrrrrrrrrr}
\hline\hline
Circle & O-C & error & Circle & O-C & error & Circle & O-C & error & Circle & O-C & error\\
\hline
15.5 & 0.00300 & 0.00022 & 20.0 & 0.0029 & 0.0003 & 22.5 & 0.0028 & 0.0004 & 34.5 & 0.00351 & 0.00018\\
37.0 & 0.00382 & 0.00022 & 39.0 & 0.00398 & 0.00023 & 44.0 & 0.0037 & 0.0003 & 56.0 & 0.00343 & 0.00016\\
68.0 & 0.00373 & 0.00026 & 72.5 & 0.00212 & 0.00029 & 75.0 & 0.0027 & 0.0004 & 80.0 & 0.00337 & 0.00017\\
82.0 & 0.00424 & 0.00027 & 84.5 & 0.00239 & 0.00024 & 99.0 & 0.0036 & 0.0003 & 101.5 & 0.00292 & 0.00020\\
108.5 & 0.00278 & 0.00028 & 111.0 & 0.00336 & 0.00016 & 113.5 & 0.00256 & 0.00020 & 115.5 & 0.0032 & 0.0004\\
118.0 & 0.00340 & 0.00028 & 120.5 & 0.00215 & 0.00017 & 127.5 & 0.0023 & 0.0003 & 139.5 & 0.00315 & 0.00022\\
151.5 & 0.0028 & 0.0003 & 154.0 & 0.00335 & 0.00024 & 161.0 & 0.00326 & 0.00017 & 161.5 & 0.00358 & 0.00023\\
166.0 & 0.00378 & 0.00023 & 173.0 & 0.00319 & 0.00016 & 175.5 & 0.00332 & 0.00016 & 178.0 & 0.00379 & 0.00016\\
204.0 & 0.00395 & 0.00024 & 206.5 & 0.0038 & 0.0003 & 209.0 & 0.00386 & 0.00020 & 216.0 & 0.0041 & 0.0004\\
218.5 & 0.00379 & 0.00018 & 221.0 & 0.002938 & 0.00027 & 223.5 & 0.0033 & 0.0003 & 226.0 & 0.00368 & 0.00019\\
228.0 & 0.00328 & 0.00020 & 228.5 & 0.0043 & 0.0006 & 230.5 & 0.00320 & 0.00016 & 240.0 & 0.00324 & 0.00016\\
252.0 & 0.00369 & 0.00020 & 254.5 & 0.0037 & 0.0005 & 261.5 & 0.0030 & 0.0003 & 264.0 & 0.0035 & 0.0003\\
273.5 & 0.0026 & 0.0003 & 276.0 & 0.00319 & 0.00027 & 285.5 & 0.00214 & 0.00016 & 765.5 & 0.0025 & 0.0006\\
825.5 & 0.00360 & 0.00021 & 837.5 & 0.00302 & 0.00022\\
\\
5429.5 & 0.0039 & 0.0012 & 5430.0 & -0.0016  & 0.0024 & 5432.5 & 0.0014 & 0.0005 & 5437.5 & 0.0030 & 0.0004\\
5439.0 & 0.0014 & 0.0006 & 5441.0 & -0.002 & 0.004 & 5441.5& 0.0034 & 0.0005 & 5442.0& 0.0025& 0.0009\\
\hline
\end{tabular}
\end{center}
\end{table}

\begin{table}
\begin{center}
\tiny
\caption{Light Curve Solutions in Different Years and Different $T_1$\label{tab4}}
\begin{tabular}{lccccccccccc}
\hline\hline
&$T_1$=5550k   &                  & $T_1$=5600K   &                  & $T_1$=5660K   &                  & $T_1$=5710K  & & overall\\
&I in 2006 & \emph{i} in 2012 & I in 2006 & \emph{i} in 2012 & I in 2006 & \emph{i} in 2012 & I in 2006 & \emph{i} in 2012 & I in 2006 & \emph{i} in 2012\\
\hline
$T_1$ & & & & & & & & & 5660(100) & 5660(100)\\
\emph{i}($^\circ$) & 77.22(10) & & 76.67(9) & & 76.81(10) & & 76.77(10) & & 76.86(23)\\
$T_2$(K) & 5443(5) & 5250(23) & 5491(4) & 5308(25) & 5544(6) & 5360(26) & 5606(7) & 5402(29) & 5521(60) & 5390(70)\\
$\Delta T=T_1-T_2$ & 107(5) & 300(23) & 109(4) & 292(25) & 116(6) & 300(26) & 104(7) & 308(29) & 109(9) & 300(30) \\
$\Omega_1=\Omega_2$ & 6.463(28) & 6.23(3) & 6.23(3) & 6.270(10) & 6.28(4) & 6.350(10) & 6.27(3) & 6.301(10) & 6.31(10) & 6.29(5)\\
q & 3.032(22) & &2.857(23) & & 2.900(27) & & 2.884(24) & & 2.91(7)\\
$L_1 / (L_1+L_2)$ & 0.2866(8) & 0.3128(19) & 0.2964(6) & 0.3214(19) & 0.2936(12) & 0.3188(18) & 0.2944(12) & 0.3183(20) & 0.293(4) & 0.318(4)\\
$\theta (rad)$ & & 0.0233(15) & & 0.025(3) & & 0.0228(20) & & 0.026(4) & & 0.0243(23)\\
$\phi (rad)$ & & 2.29(10) & & 2.27(10) & &2.22(10) & &2.17(13) & & 2.23(14)\\
$r_{spot}(rad)$ & & 0.737(22) & & 0.732(4)& & 0.71(6) & &0.82(12) & & 0.75(5)\\
$T_s/T_*$ & & 1.65(4) & & 1.654(17) & & 1.70(9) & &1.54(18) & & 1.63(21)\\
$\sigma=\overline{ \sum (O-C)^2 }$ & 0.004 & 0.005 & 0.004 & 0.005 & 0.004 & 0.005 & 0.004 & 0.005 \\

\hline

\hline
\end{tabular}
\end{center}
\end{table}





\begin{thebibliography}{}
\bibitem[Barnes et al.(2005)]{2005MNRAS.356.1501B} Barnes, J.~R., Collier 
Cameron, A., Lister, T.~A., Pointer, G.~R., 
\& Still, M.~D.\ 2005, \mnras, 356, 1501 
\bibitem[Berdyugina(2005)]{2005LRSP....2....8B} Berdyugina, S.~V.\ 2005, 
Living Reviews in Solar Physics, 2, 8 
\bibitem[Frasca et 
al.(2008)]{2008A&A...479..557F} Frasca, A., Biazzo, K., Ta{\c s}, G., Evren, S., \& Lanzafame, A.~C.\ 2008, \aap, 479, 557 
\bibitem[Johnson(1966)]{1966ARA&A...4..193J} Johnson, H.~L.\ 1966, \araa, 4, 193
\bibitem[Kron(1947)]{1947PASP...59..261K} Kron, G.~E.\ 1947, \pasp, 59, 261 
\bibitem[Li et al.(2014)]{2014AJ....147...98L} Li, K., Qian, S.-B., Hu, 
S.-M., \& He, J.-J.\ 2014, \aj, 147, 98
\bibitem[Liu et al.(2012)]{2012PASJ...64...48L} Liu, L., Qian, S.-B., He, 
J.-J., et al.\ 2012, \pasj, 64, 48 
\bibitem[Lucy(1976)]{1976ApJ...205..208L} Lucy, L.~B.\ 1976, \apj, 205, 208 
\bibitem[Luo et al.(2015)]{2015AJ....150...70L} Luo, C.~Q., Zhang, X.~B., 
Deng, L., Wang, K., \& Luo, Y.\ 2015, \aj, 150, 70 
\bibitem[Mullan(1975)]{1975ApJ...198..563M} Mullan, D.~J.\ 1975, \apj, 198, 
563
\bibitem[Neff et al.(1995)]{1995ApJ...452..879N} Neff, J.~E., O'Neal, D., 
\& Saar, S.~H.\ 1995, \apj, 452, 879 
\bibitem[Niarchos et 
al.(1996)]{1996A&AS..117..105N} Niarchos, P.~G., Hoffmann, M., \& Duerbeck, H.~W.\ 1996, \aaps, 117, 105
\bibitem[O'Connell(1951)]{1951PRCO....2...85O} O'Connell, D.~J.~K.\ 1951, 
Publications of the Riverview College Observatory, 2, 85 
\bibitem[O'Neal et al.(1996)]{1996ApJ...463..766O} O'Neal, D., Saar, S.~H., 
\& Neff, J.~E.\ 1996, \apj, 463, 766 
\bibitem[Qian et al.(2012)]{2012MNRAS.423.3646Q} Qian, S.-B., Zhang, J., 
Zhu, L.-Y., et al.\ 2012, \mnras, 423, 3646
\bibitem[Qian et al.(2014)]{2014ApJS..212....4Q} Qian, S.-B., Wang, J.-J., 
Zhu, L.-Y., et al.\ 2014, \apjs, 212, 4 
\bibitem[Qian et al.(2015)]{2015AJ....149...38Q} Qian, S.-B., Essam, A., 
Wang, J.-J., et al.\ 2015, \aj, 149, 38 
\bibitem[Ruci{\'n}ski(1969)]{1969AcA....19..245R} Ruci{\'n}ski, S.~M.\ 
1969, \actaa, 19, 245 
\bibitem[Sterken(2005)]{2005ASPC..335....3S} Sterken, C.\ 2005, The 
Light-Time Effect in Astrophysics: Causes and cures of the O-C diagram, 
335, 3 
\bibitem[Strassmeier(2009)]{2009A&ARv..17..251S} Strassmeier, K.~G.\ 2009, \aapr, 17, 251 
\bibitem[Szyma{\'n}ski et al.(2010)]{2010AcA....60..295S} Szyma{\'n}ski, 
M.~K., Udalski, A., Soszy{\'n}ski, I., et al.\ 2010, \actaa, 60, 295
\bibitem[Tran et al.(2013)]{2013ApJ...774...81T} Tran, K., Levine, A., 
Rappaport, S., et al.\ 2013, \apj, 774, 81
\bibitem[van Hamme(1993)]{1993AJ....106.2096V} van Hamme, W.\ 1993, \aj, 
106, 2096 
\bibitem[Vilhu(1981)]{1981Ap&SS..78..401V} Vilhu, O.\ 1981, \apss, 78, 401

\bibitem[Wilson 
\& Devinney(1971)]{1971ApJ...166..605W} Wilson, R.~E., \& Devinney, E.~J.\ 1971, \apj, 166, 605 
\bibitem[Wilson(1979)]{1979ApJ...234.1054W} Wilson, R.~E.\ 1979, \apj, 234, 
1054 
\bibitem[Wilson(1990)]{1990ApJ...356..613W} Wilson, R.~E.\ 1990, \apj, 356, 
613 
\bibitem[Wilson(1994)]{1994PASP..106..921W} Wilson, R.~E.\ 1994, \pasp, 
106, 921 
\bibitem[Wilson(2006)]{2006ASPC..349...71W} Wilson, R.~E.\ 2006, 
Astrophysics of Variable Stars, 349, 71 

\bibitem[Zacharias et al.(2013)]{2013AJ....145...44Z} Zacharias, N., Finch, 
C.~T., Girard, T.~M., et al.\ 2013, \aj, 145, 44 








 








\label{bibliography}



\end{thebibliography}
\end{document}